\documentclass[a4paper,aps,twocolumn]{revtex4-2}
\usepackage{amsmath,setspace,graphicx,xcolor,textcomp}
\usepackage{epstopdf}
\usepackage[T1]{fontenc}
\usepackage{hyperref}

\newcommand{\revisiontwo}[1]{\textcolor{black}{#1}}
\newcommand{\revisionthree}[1]{\textcolor{black}{#1}}

\usepackage{url}

\begin{document}
\title{Moo-ving mountains: grazing agents drive terracette formation on steep hillslopes}

\author{Benjamin Seleb}
\affiliation{Interdisciplinary Graduate Program in Quantitative Biosciences, Georgia Institute of Technology, Atlanta, GA, United States}

\author{Louis Gonz\'alez}
\affiliation{School of Chemical and Biomolecular Engineering, Georgia Institute of Technology, Atlanta, GA, United States}
\affiliation{BioFrontiers Institute, University of Colorado, Boulder, CO, United States}
\affiliation{Department of Chemical and Biological Engineering, University of Colorado, Boulder, CO, United States}

\author{Atanu Chatterjee}
\affiliation{School of Chemical and Biomolecular Engineering, Georgia Institute of Technology, Atlanta, GA, United States}
\affiliation{BioFrontiers Institute, University of Colorado, Boulder, CO, United States}
\affiliation{Department of Chemical and Biological Engineering, University of Colorado, Boulder, CO, United States}

\author{Saad Bhamla}
\email{saadb@chbe.gatech.edu}
\affiliation{School of Chemical and Biomolecular Engineering, Georgia Institute of Technology, Atlanta, GA, United States}
\affiliation{BioFrontiers Institute, University of Colorado, Boulder, CO, United States}
\affiliation{Department of Chemical and Biological Engineering, University of Colorado, Boulder, CO, United States}

\begin{abstract}
Terracettes, striking, step‑like landforms that stripe steep, vegetated hillslopes, have puzzled scientists for more than a century. Competing hypotheses invoke either slow mass‑wasting (gravity-driven soil flow) or the relentless trampling of grazing animals, yet no mechanistic model has linked hoof‑scale behavior to landscape‑scale form. Here we bridge that gap with an active‑walker model in which ungulates are represented as stochastic foragers moving on an erodible slope. Each agent weighs the energetic cost of climbing against the benefit of fresh forage; every hoof‑fall compacts soil and lowers local biomass, subtly reshaping the energy landscape that guides subsequent steps. Over time, these stigmergic feedbacks concentrate traffic along cross‑slope paths that coalesce into periodic tread‑and‑riser bands, morphologically analogous to natural terracettes. Our model illustrates how local foraging rules governing movement and substrate feedback can self‑organize into large‑scale topographic patterns, highlighting the wider role of decentralized biological processes in sculpting terrestrial landscapes. 
\end{abstract}

\maketitle

\section*{Introduction}
Natural landscapes are replete with intriguing structures that arise from the reciprocal interaction of animals and their physical environment. Over a century ago, Darwin~\cite{darwin1892formation} described how earthworms leave persistent and profound imprints on the soil they inhabit, offering an early glimpse into what is now known as biogeomorphology~\cite{vilesbiogeomorphology}. Subsequent research has since broadened our understanding of how diverse ecosystem engineers—from burrowing animals that bioturbate soil \cite{germain2025hidden} to large termite colonies that construct massive mounds~\cite{funch2015termite, ocko2019morphogenesis}, can reshape landscapes across a variety of habitats and time scales.

Despite clear evidence of these biogenic forces in shaping landscapes, traditional geoscientists have historically emphasized geophysical processes, such as tectonic uplift and mass wasting, over biological ones, a bias previously referred to as ``geophysical orthodoxy''~\cite{rice2021so}. These perspectives tend to overshadow the potential for local animal activities to generate large-scale geomorphic effects, often deeming such contributions as secondary or superficial.

However, recent studies increasingly challenge this assumption. A recent global assessment estimates that the annual geomorphic energy contributed by wild animals may exceed that of thousands of extreme
flood events \cite{harvey2025global}. Additionally, repeated local activities of animals often foster spatiotemporal feedback, where incremental actions accumulate into self-reinforcing dynamics that amplify their overall geomorphic impact~\cite{murray2008biomorphodynamics,francis2009perspectives, polvi2012beaver, stallins2006geomorphology, perry1995self}. 

These feedbacks do more than amplify impact; in many ecological systems, they also impose order. Continuous organism–environment interactions can give rise to large-scale spatial patterns—regularities often described using principles from statistical physics and complex systems theory~\cite{tarnita2024self, tarnita2018ecology, pringle2017spatial}. Classic examples include patterns and bands in vegetation \cite{rietkerk2004self}, hexagonal spacing of animal nests and territories \cite{tarnita2018ecology, barlow1974hexagonal}, and other spatial structures that emerge without centralized control~\cite{holling1996self, pringle2017spatial}.

\begin{figure}[ht!]
\includegraphics[width=0.48\textwidth]{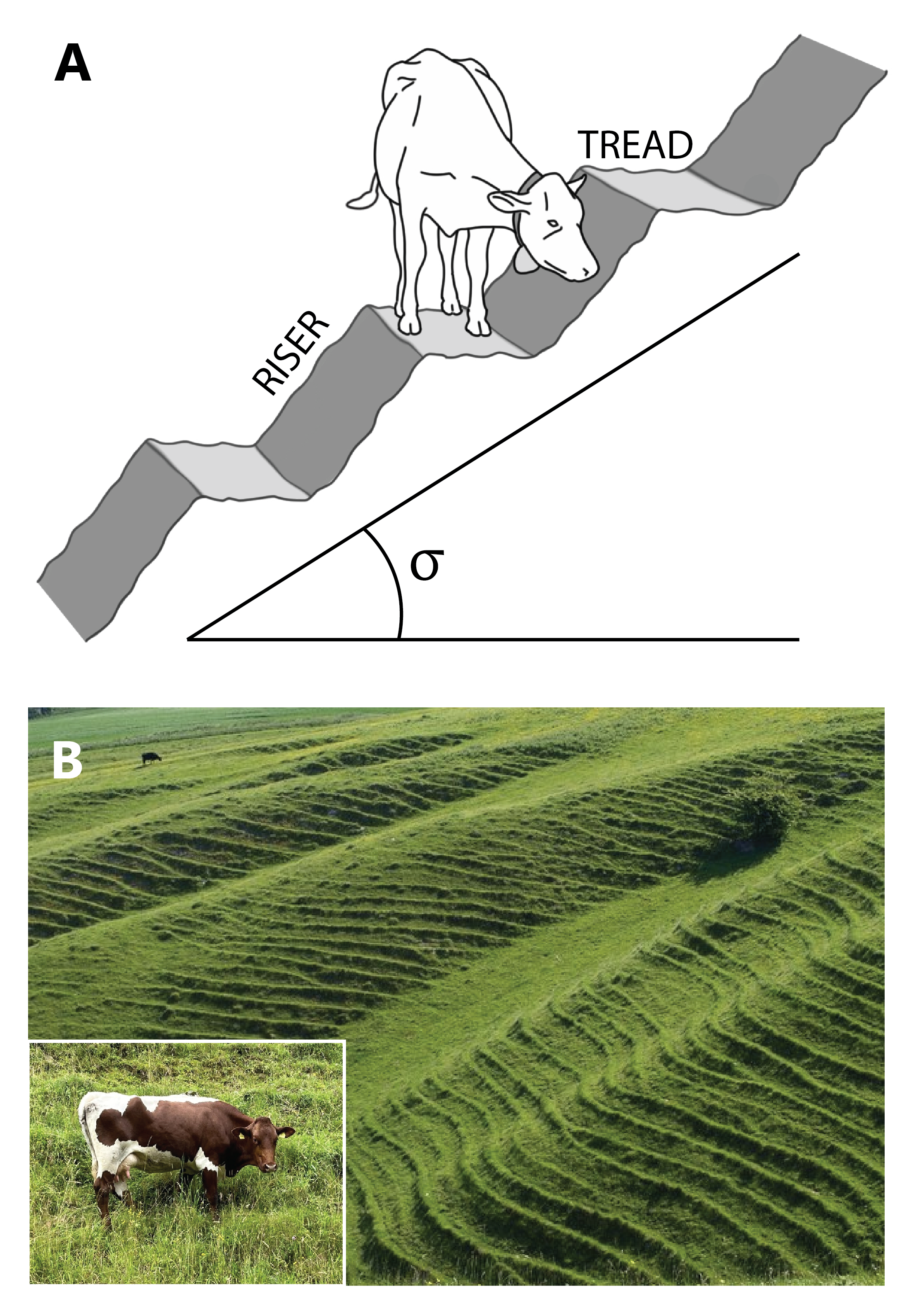}
\caption{\textbf{Terracette morphology.} Terracettes are recognized as periodic landscape patterns that exhibit long, contour-following pathways along hillslopes. Each terracette step consists of a flat tread and a sloped riser, with adjacent steps often interconnected by shorter sloped ramps. (A) A diagram illustrating a hillside with major slope angle $\sigma$, featuring terraced steps made of treads and risers. (B) A particularly striking example of terracettes in Wiltshire, England~\cite{morganshill}.}
\label{fig:diagram} 
 \vspace{-15pt}
\end{figure}

Despite their striking regularity, \textit{terracettes}--closely spaced, terrace-like steps that contour vegetated hillslopes--have not been examined through this lens. Found in diverse environments from arid grasslands to alpine meadows, terracettes have long been a subject of debate regarding their origins. Often referred to as ``cow paths'' or ``sheep trails'', these features have long been colloquially attributed to the activity of livestock and wild ungulates (hoofed mammals), which are commonly observed traversing the flatter treads while grazing on the intervening risers (see Figure~\ref{fig:diagram}). 

Indeed, ungulates are well-established as geomorphic agents~\cite{trimble1995cow}, and multiple studies report strong correlations between terracette morphology and ungulate activity~\cite{brice1958origin,watanabe1994soil,hartwig2025gully}. In one case, terracettes were observed to form within weeks of intensive sheep grazing~\cite{higgins1982}, while another linked their geometric properties to animal morphology~\cite{howard1987dimensions}. Despite these observations, the precise mechanism by which grazing could produce such regular, periodic landforms remains poorly understood.

A central objection to a biogenic origin is the apparent mismatch between the erratic, meandering movements of grazing animals and the highly ordered, repeating structure of terracettes~\cite{vincent1980terracette}. Consequently, the prevailing orthodoxy has attributed terracette formation to geophysical processes such as soil creep, slumping, or periglacial activity--relegating ungulates to a largely superficial role~\cite{odum1922om, buckhouse1981caused, bielecki2002origin, kuck2002terracettes, Ehlers2022}.

We argue that this orthodoxy is misleading. Here, we propose an agent-based model that represents ungulates as random walkers on an erodible terrain. Similar \textit{active walker} models have been used to study trail formation in both animals and humans~\cite{schweitzer1997active, helbing1997active, gilks2009mountain}.  
Guided by local foraging decisions that balance energy expenditure in navigating slopes, these agents gradually reshape the terrain through repeated trampling and soil compaction. This altered terrain, in turn, biases their subsequent movement decisions via a cycle of biogeomorphic feedback, creating conditions ripe for self-organization. Over time, this feedback gives rise to terrace-like steps, demonstrating that terracettes may arise independently from the self-organized interactions of organisms constrained by energetic landscapes.

\color{black}

\section*{Model}
We define a sloped landscape, represented by the heightmap, $\mathcal{H}\left(x,y,t\right)$ on which ungulates graze as active agents. This landscape is defined over a square domain of side length $s$, initialized with a linear slope along the $y$-direction. Specifically, at $t=0$, the terrain follows $\mathcal{H}\left(x,y,t=0\right)=y\tan\sigma$, where $\sigma$ is the pitch angle and the positive $y$-direction points uphill. A periodic boundary condition is imposed such that agents crossing a boundary reappear on the opposite side. This choice eliminates boundary artifacts and reduces computational cost, but it introduces additional modeling considerations (e.g., resource regeneration on a finite landscape).

In addition to the physical landscape, we define a resource map $\mathcal{R}\left(x,y,t\right)$, characterized by a linear regrowth rate, $\mu$. The resource map is normalized to range from 0 (depleted) to 1 (abundant) and is subject to the same periodic boundaries. Although simplified, this representation of vegetation growth can be interpreted as a shared, stigmergic ``working memory''~\cite{bailey1996mechanisms}: depleted patches persist as traces of recent traffic (even as agents wrap across the periodic boundary), and the regrowth rate $\mu$ controls how quickly those traces fade. In this sense, $\mathcal{R}$ encodes perceived forage availability rather than literal biomass.

Each agent occupies a position $\left(x,y,z\right)$, where $z = \mathcal{H}\left(x,y,t\right)$ denotes the local elevation at time $t$. Agents explore the environment in discrete time steps, moving along a sequence of straight paths of length $l$, each in a new direction $\phi$---a standard approach in random walk models~\cite{codling2008random,farnsworth1999grazers,smouse2010stochastic}. In many foraging models, $l$ is drawn sequentially from a probability density function, often heavy-tailed (L{\'e}vy flight)~\cite{zhao2016understanding, romero2023grazing}. This approach reproduces large-scale patterns of movement, though it does not explicitly represent the local interactions that generate them. On uniform terrains this abstraction is often adequate, where the underlying drivers of direction change are often indistinguishable from noise. On sloped terrains, however, the locomotion costs are direction-dependent~\cite{shepard2013energy}, so directional decisions cannot be sampled from an invariant (isotropic) dispersal kernel. 

We instead define the path length $l$ as a short, fixed decision horizon at which agents reassess their direction of motion based on the local energy landscape. An agent's field of view is split into $B$ azimuthal bins, and at each time step, it computes a travel cost for each prospective direction $\phi_i$ $\left(i\in\left[1,\ldots,B\right]\right)$. By weighting direction choice with this cost, the model yields a biased random walk, with movement dynamics shaped by informed forager–landscape interactions~\cite{smouse2010stochastic, fronhofer2013random}. 

\begin{figure}[t!]
\includegraphics[width=.45\textwidth]{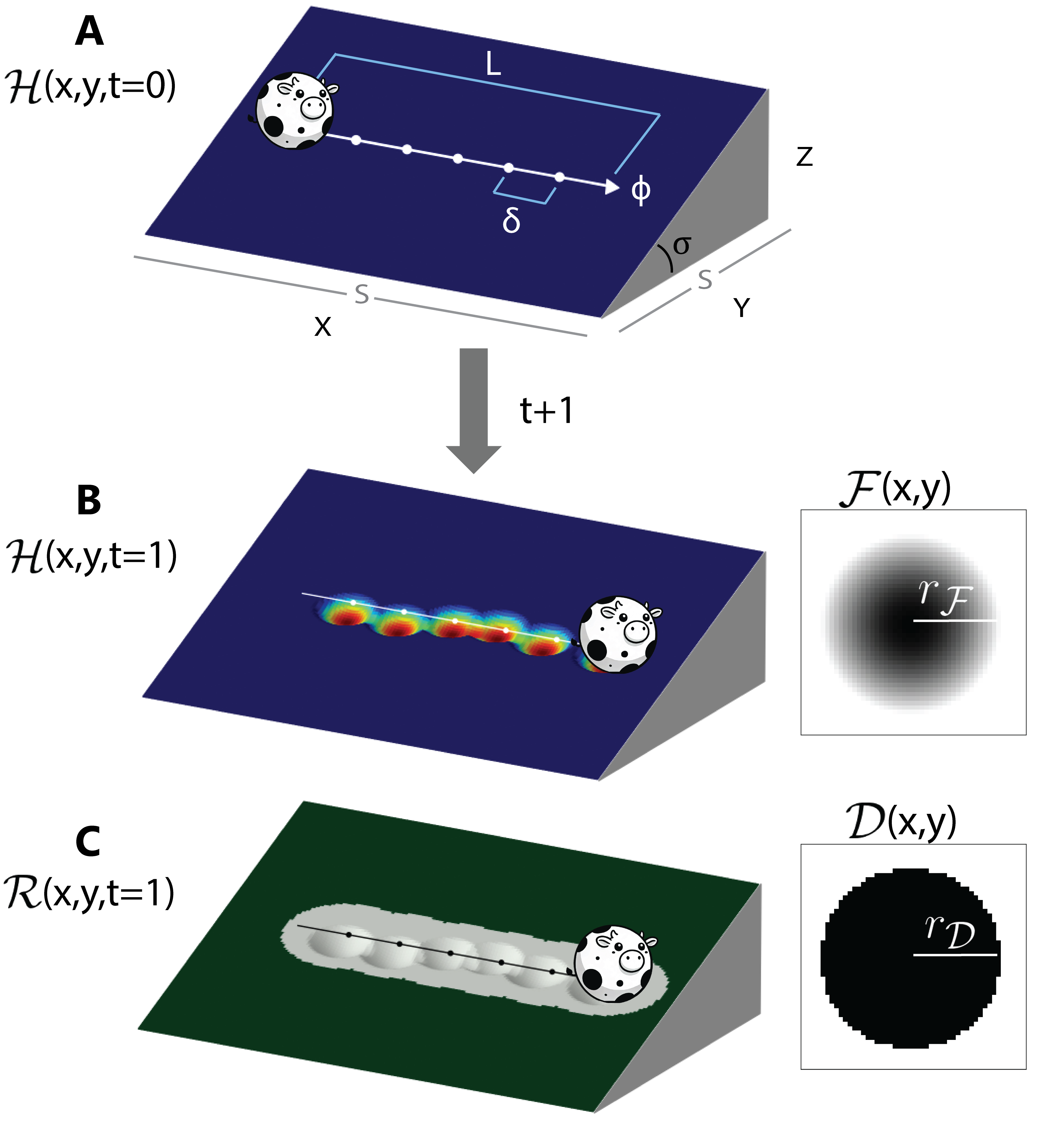}
\caption{\label{fig:model} \textbf{Agent behavior over a single timestep.} (A) An agent (spherical cow) is initialized on the terrain heightmap $\mathcal{H}(x,y,t=0)$. During the subsequent timestep ($t = 1$, arbitrary units), the agent travels along a path of length $l$ in direction $\phi$, making discrete footsteps at an interval $\delta$. 
(B) At each footstep location, the agent applies a circular erosion kernel $\mathcal{F}$ of radius $r_\mathcal{F}$, representing a footprint that contributes to the updated heightmap $\mathcal{H}(x,y,t=1)$. 
(C) Simultaneously, a uniform depletion kernel $\mathcal{D}$ of radius $r_\mathcal{D}$ is applied to represent foraging reach. The resulting updated resource map $\mathcal{R}(x,y,t=1)$ is displayed and projected onto the surface.}
 \vspace{-15pt}
\end{figure}

After choosing a direction, the agent subdivides the path into $n$ discrete ``footsteps'' of length $\delta$, such that $n = l/\delta$, as shown in Figure~\ref{fig:model}A. The footstep increment for the agent is given by, 
\begin{equation}
\begin{gathered}
x_{k+1}\left(t\right)=x_k\left(t\right)+\delta\sin\phi_i\left(t\right)\\
y_{k+1}\left(t\right)=y_k\left(t\right)+\delta\cos\phi_i\left(t\right)
\end{gathered}
\end{equation}

As each agent $a\in\left[1,\ldots,N\right]$ moves, every footstep depletes local resources and tramples the soil (Figure~\ref{fig:model}B-C). We define these environmental modifications using landscaping functions, akin to those in active walker models \cite{lam2005active}. 
Specifically, resources are consumed within a radius $r_\mathcal{D}$ of each footstep, following a uniform depletion kernel $\mathcal{D}(x, y)$--a proxy for the agent’s neck reach or foraging radius, within which vegetation is consumed.
Simultaneously, the terrain is modified by a smooth, circular footprint kernel $\mathcal{F}(x, y)$, which is intended to represent localized trampling pressure. 
To reflect natural variation in foot placement, each footprint location $(x^a_k, y^a_k)$ is randomly perturbed by a small offset $(v_x, v_y)$, so that the actual depression is centered at $(\widetilde{x}^a_k, \widetilde{y}^a_k) = (x^a_k + v_x, y^a_k + v_y)$, where $v_x$ and $v_y$ are independently sampled from the interval $[-\eta, \eta]$. For full definitions of the landscaping kernels $\mathcal{D}$ and $\mathcal{F}$, see Figure~S1 and the accompanying Supplementary Information.

\begin{figure*}
\centering
\includegraphics[width=0.95\textwidth]{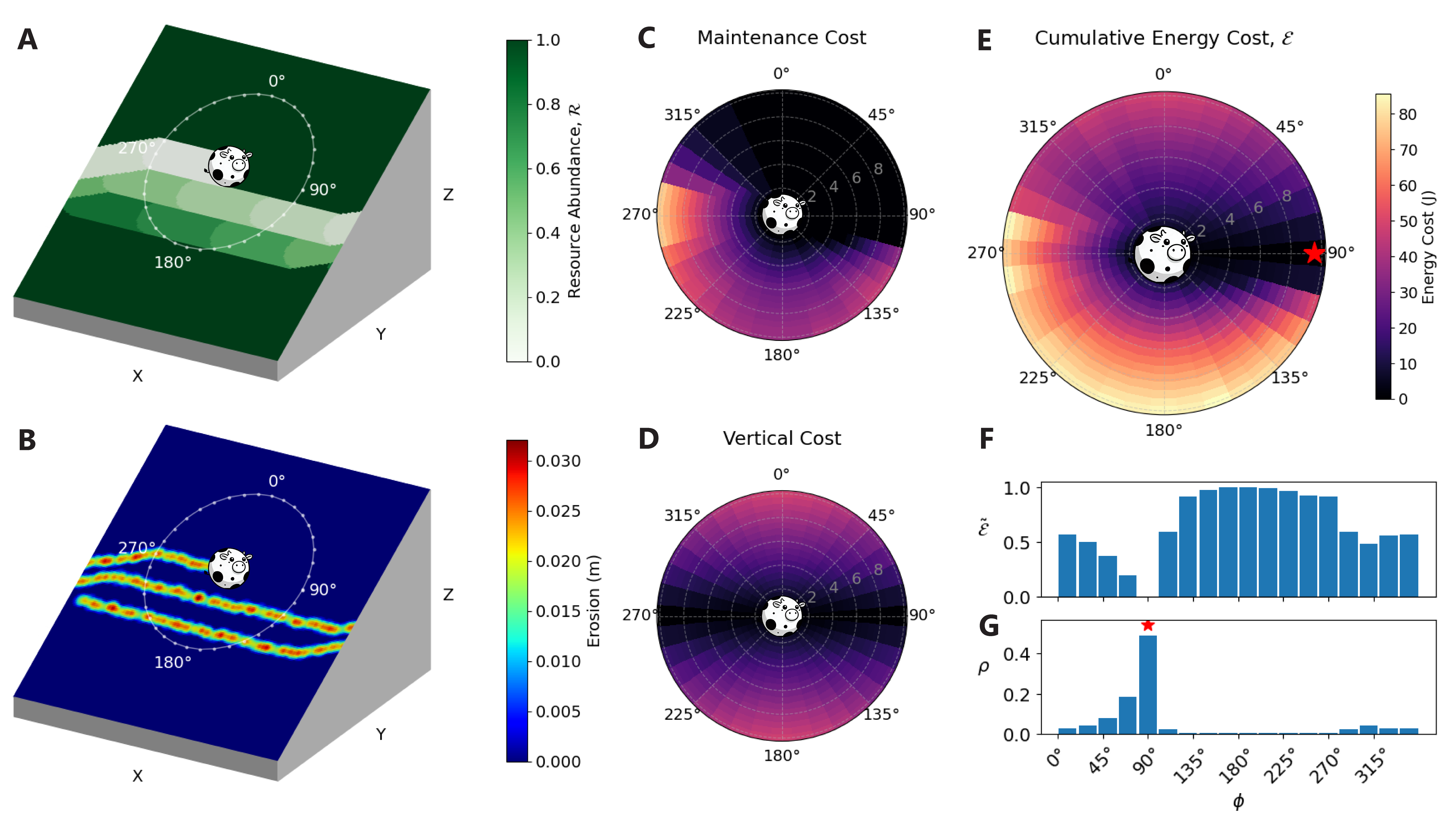} 
\caption{\textbf{Agent-centric energy landscape on a sloped terrain.} (A) Resource map $\mathcal{R}$, illustrating the trail of depleted resources along the agent's path, which wraps around the periodic boundary condition. As resources replenish, the trail of depletion gradually fades. Potential future positions are marked on a circular ring at a distance $l$ from the agent. (B) Terrain heightmap $\mathcal{H}$, depicting the surface depressions in the terrain resulting from the movement of the same agent. (C) The maintenance cost map represents the energy required for the agent to reach any point within a radius $l=10$m from its current position. The highest costs are directly behind the agent, reflecting where resource depletion is most recent. Note that panels C, D, and E share the same color bar. (D) The vertical (climbing) cost map indicates the additional energy required to perform vertical locomotion (uphill and downhill). (E) Total energy cost, which combines both maintenance and climbing costs. The cost increases sharply in areas with significant elevation gradients and/or resource depletion. (F) Normalized energy cost $\widetilde{\mathcal{E}}$: The cost associated with moving the full distance $l=10$ m in each direction, normalized for comparison. (G) The probability distribution $\rho(\phi)$, computed using the exponential of the scaled energy cost $\exp\left(-\beta~\widetilde{\mathcal{E}}(\phi_i)\right)$. The most probable movement direction (red star) corresponds to the direction with the lowest total energy cost (as seen in panel E).}
\label{fig:terracette}
\end{figure*}

For each successive time step $t+1$, we update both the resource map $\mathcal{R}\left(x,y,t+1\right)$ and the terrain height map $\mathcal{H}\left(x,y,t+1\right)$ based on the cumulative effects of resource depletion and soil erosion from all agents' footsteps. Specifically:

\vspace{-10pt}
\begin{equation}
    \begin{gathered}
        \mathcal{R}\left(x,y,t+1\right) = \left(\mathcal{R}\left(x,y,t\right) + \mu\right) -
    \sum_{a=1}^{N}\sum_{k=1}^{n} \mathcal{D}\left(x^a_k,y^a_k\right) \\
    \mathcal{H}\left(x,y,t+1\right) = \mathcal{H}\left(x,y,t\right) - \sum_{a=1}^{N}\sum_{k=1}^{n}\mathcal{F}\left(\widetilde{x}^a_k,\widetilde{y}^a_k\right)
    \end{gathered}
\end{equation}

where $N$ is the total number of agents, $n$ is the number of footsteps along each path, and $\mu$ is the rate at which resources replenish. After updating $\mathcal{R}\left(x,y,t+1\right)$, we clip its values to the interval $\left[0,1\right]$ to ensure that resources never become negative or exceed full capacity. As illustrated in Figure~\ref{fig:terracette}A-B, the agents' footprints continuously reshape both the resource distribution and the terrain.

Each footstep along an agent's path incurs an energetic cost determined by both the terrain's local slope and the resource abundance. We decompose this cost into two primary components: a baseline maintenance cost and a vertical travel cost. Formally, the stepwise cost for the $k$-th footstep in direction $\phi_i$ is given by,
\begin{align}
e_k(\phi_i) = &\underbrace{\omega_h\delta \left(1-\mathcal{R}\left(x_{k+1}, y_{k+1}\right)\right)}_{\text{maintenance}} \\
            &+ \underbrace{\omega_v\left|\mathcal{H}\left(x_{k+1}, y_{k+1}\right)-\mathcal{H}\left(x_k, y_k\right) \right|}_{\text{vertical}}
\end{align}

where $\delta$ is the footstep length.

The maintenance cost captures baseline energetic expenditures associated with horizontal locomotion, chewing, and metabolism~\cite{brosh2006energy}, with $\omega_h$ reflecting the energy cost per meter traveled while grazing (see Figure~\ref{fig:terracette}C). In areas where resources are plentiful $\left(\mathcal{R}\left(x,y\right)\approx 1\right)$, nutritional intake can partially offset this cost, whereas limited resources $\left(\mathcal{R}\left(x,y\right)< 1\right)$ lead to higher net energy expenditures.

Movement on sloped terrain introduces an additional vertical travel cost, representing the energy needed to lift or stabilize the animal's body mass~\cite{shepard2013energy,taylor1972running}. This cost is computed by multiplying the energy cost per unit of vertical travel, $\omega_v$, by the absolute elevation change over each step (Figure~\ref{fig:terracette}D). While downhill travel can theoretically allow animals to recover some energy, it often entails additional costs due to braking and stabilization forces~\cite{taylor1972running,minetti2002energy,birn2014scaling,carnahan2021quantifying}.

Finally, the cumulative cost $\mathcal{E}(\phi_i)$ incurred by the agent while traveling a distance $l$ in direction $\phi_i$ from its current position is obtained by summing the stepwise costs,

\vspace{-10pt}
\begin{equation}
\mathcal{E}(\phi_i) = \sum_{k = 1}^{n} e_k(\phi_i)
\end{equation}

Each azimuthal orientation $\phi_i$ represents a possible state for the agent, each with an associated energy cost $\mathcal{E}(\phi_i)$, as illustrated in Figure~\ref{fig:terracette}E. On inclined terrain ($\sigma\neq0$), directions that minimize elevation gain become more energetically favorable. To model the agent's decision among these orientations, we use a probabilistic stepping rule:

\vspace{-10pt}
\begin{equation} 
    \rho(\phi_i) \propto \exp\left(-\beta~\widetilde{\mathcal{E}}(\phi_i)\right).
\end{equation}

where $\rho\left(\phi_i\right)$ is the probability of moving distance $l$ in direction $\phi_i$. This formulation resembles a Boltzmann active walk~\cite{pochy1993boltzmann,lam2005active}, with $\widetilde{\mathcal{E}}(\phi_i)$ denoting a normalized travel cost, and $\beta>0$ serving as an inverse temperature term that governs the agent's sensitivity to energetic differences. In the limit $\beta=0$, $\rho\left(\phi\right)$ reduces to a uniform random walk. As $\beta$ increases, the most probable movement direction corresponds to the lowest total energy cost (Figure~\ref{fig:terracette}F-G). 

\begin{figure}[t!]
\includegraphics[width=0.48\textwidth]{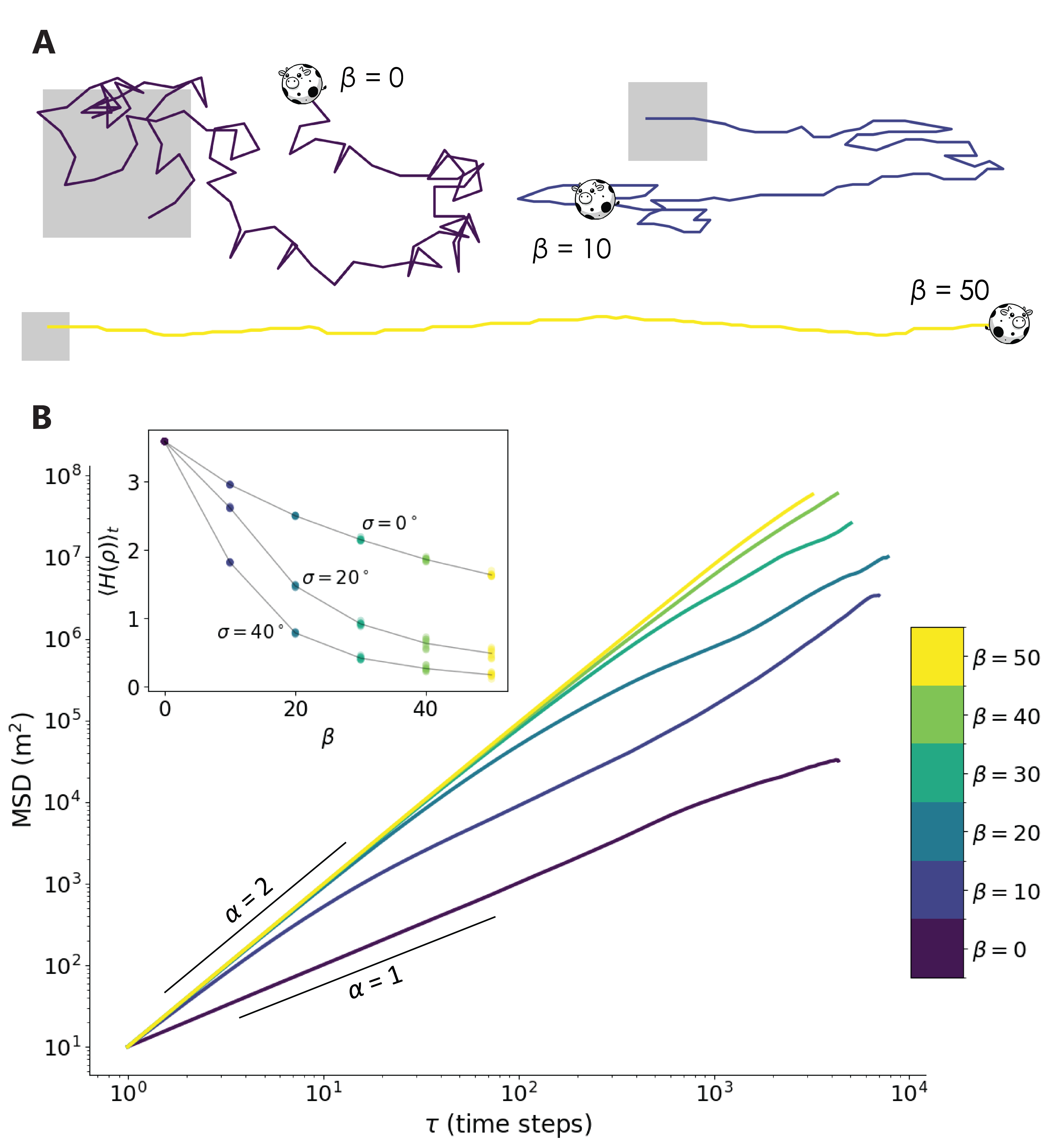}
\caption{{\textbf{Energetic sensitivity shapes random walk dynamics.} 
(A) Representative trajectories (100 steps) of agents walking over a slope of $\sigma = 40^\circ$, with increasing sensitivity to energetic cost ($\beta$). Paths are unwrapped from the periodic boundary condition to reveal cumulative displacement, and the underlying surface (size $s$) is indicated by the gray square. (B) Mean square displacement (MSD) as a function of lag time ($\tau$), averaged across an ensemble of 10 simulations per $\beta$ (also with $\sigma = 40^\circ$).
\textbf{Inset:} Time-averaged Shannon entropy of $\rho(\phi)$ shown for multiple slopes ($\sigma = 0^\circ$, $20^\circ$, and $40^\circ$).
}}
\label{fig:walkfig} 
\end{figure}

By systematically varying $\beta$, we reveal distinct dynamical regimes -- ranging from diffuse wandering to strongly self-reinforcing path formation -- that illustrate how ungulates adapt their grazing strategies under changing energy constraints. As shown in Figure~\ref{fig:walkfig}A, movement trajectories shift from tortuous paths at low $\beta$ to persistent motion at high $\beta$. This transition is quantified by the scaling exponent of the mean-square displacement, which changes from diffusive $\left(\alpha \approx 1\right)$ to ballistic $\left(\alpha \approx 2\right)$ as $\beta$ increases (Figure~\ref{fig:walkfig}B). Thus, even with a fixed path length, the model is capable of producing superdiffusive trajectories. From an information-theoretic perspective, the Boltzmann rule can be viewed as the distribution that maximizes Shannon (Gibbs) entropy subject to an energy cost constraint. As both $\beta$ and $\sigma$ increase, the orientation distributions, $\rho\left(\phi\right)$ exhibit lower entropy, reflecting increasingly selective movement choices driven by heightened energetic sensitivity and steeper terrain gradients (Figure~\ref{fig:walkfig}B, inset). 

In all simulations, $N=3$ agents were placed randomly on an untouched terrain $\mathcal{H}$ of size $s=50$ m, coupled with an initially abundant resource map $\mathcal{R}$. Other simulation parameters were selected to lie within the regime where terracette-like banding emerges. The influence of several key parameters is explored in supplemental analyses (Figures~S2–S3), while Table~S1 provides a complete list of parameter values together with their reasoning. 

\begin{figure*}[t!]
\centering
\includegraphics[width=\textwidth]{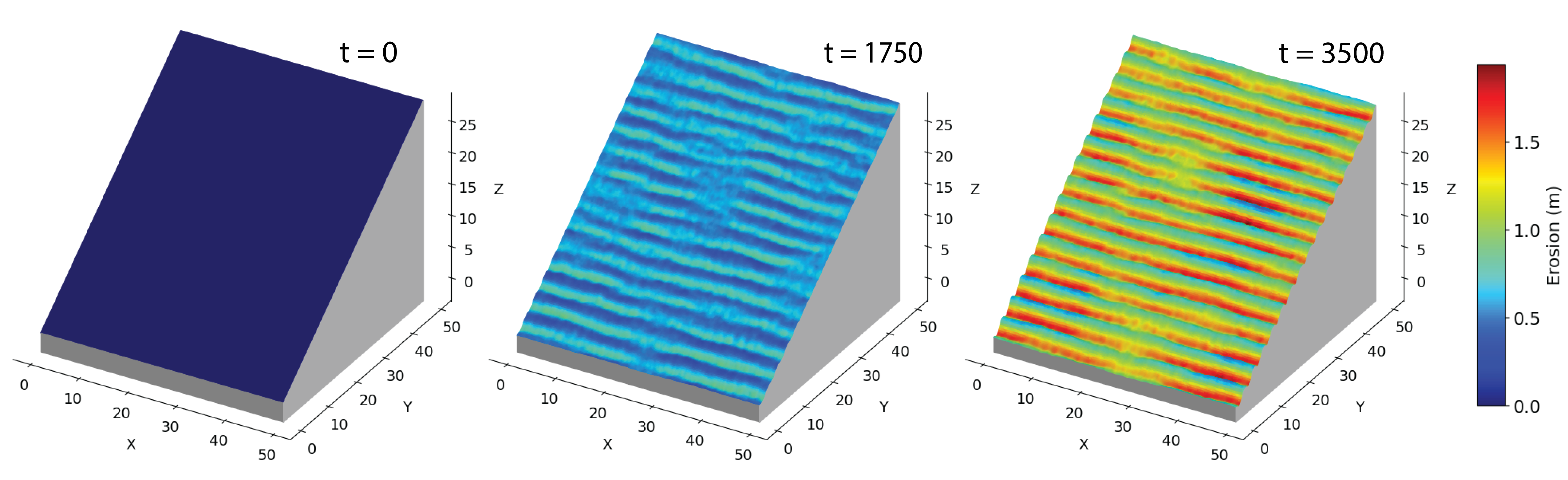} 
\caption{\textbf{The simulated process of terracette formation.} By simultaneously sensing and modifying the simulated environment, agents gradually form terracettes, with increasing depth and regularity over successive timesteps. The current simulation involves $N=3$ agents, with parameters set at $\beta = 50$, $\sigma = 30^\circ$. Please refer to Movies S1-S2 for a more detailed visualization of the erosion process.}
\label{fig:process}
\end{figure*}
 
\section*{Results}
\subsection*{Self-organized pattern formation}
Figure~\ref{fig:process} presents snapshots of a single simulation on a steep slope $\left(\sigma=30^{\circ}\right)$. At $t=0$, the terrain is nearly smooth and erodes rather uniformly. As time progresses, however, agents' stochastic movements lead to uneven traversals; some areas experience higher frequencies of trampling, resulting in greater compaction and thus increased surface deviation. These deviations deepen through a nonlinear feedback loop, where repeated trampling further stabilizes and accentuates the same tracks, ultimately forming visible trails of compaction that mark frequently used paths. By $t\approx1750$, these horizontal trails transition the landscape from isotropic erosion to anisotropic networks of well-defined paths. Similar behavior has been observed in other active walker models~\cite{lam1995active}, in which agents become trapped in their own troughs, further intensifying the grooves. Ultimately, by $t=3500$, ongoing reciprocal interaction between agents and terrain gives rise to periodic surface morphology that strongly resembles terracettes.

\subsection*{Resource regeneration on a finite landscape}

\revisionthree{Because the domain is finite and periodically wrapped, even a few agents impose a grazing pressure much higher than what is realistic. As such, the balance between resource regrowth and depletion becomes a critical consideration. While regrowth is quite straightforward, controlled by $\mu$ and $r_\mathcal{F}$, the rate of depletion is influenced by several parameters controlling agent movement, including $\eta, r_\mathcal{D}, \delta, l$ and $\gamma$. To examine this balance more simply, we consider the relative timescales of resource recovery and agent revisitation. We approximate the recovery timescale as the inverse of the regrowth rate, $1/\mu$, and the revisit timescale as $s/l$, where \(s\) is the domain side length and \(l\) is the step length per time step.} 

\begin{figure*}[t!]
\centering
\includegraphics[width=0.98\textwidth]{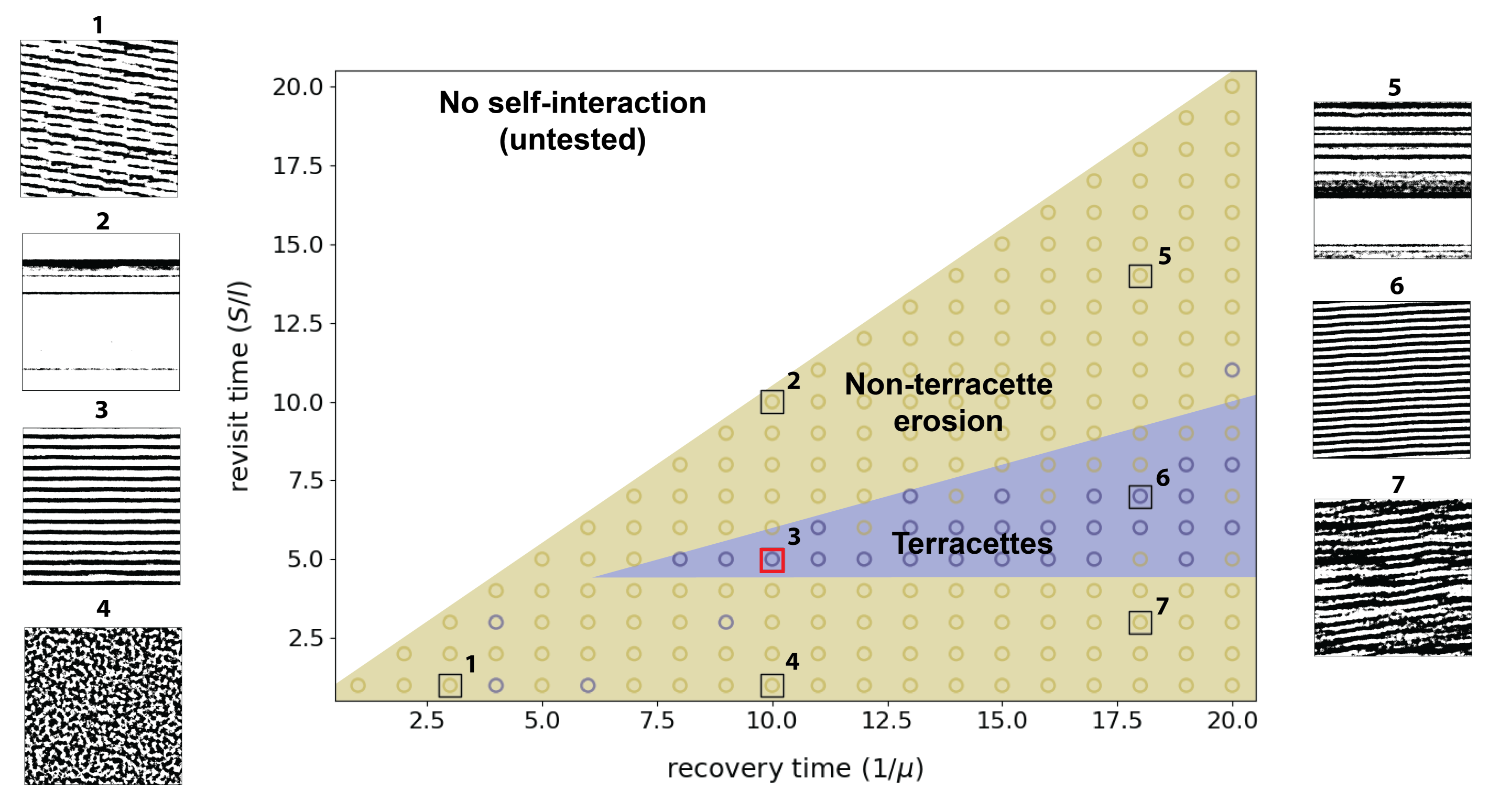}
\caption{
Revisit time vs. recovery time. Axes show recovery time $1/\mu$ (time for a fully depleted patch to regrow) and revisit time $s/l$ (time, in simulation steps, for an agent with step length $l$ to traverse one domain side of length $s$ under periodic boundaries). Each circle is a single simulation with one agent ($N=1$) with $\beta=50$ and slope angle $\sigma=35^\circ$. Representative snapshots of deformation maps are shown for a subset of these simulations. Simulations are separated using an autocorrelation threshold ($\bar{R}(\lambda)>0.75$) into approximate, schematically indicated regimes representing “terracettes” (blue wedge) and “non-terracette erosion” (yellow). Because the model dynamics are stochastic, regime boundaries are not perfectly discrete. The white region above the diagonal ($1/\mu<s/l$) was not examined because any resource marks fade before agents can revisit them, precluding self-interaction. Even below the diagonal, self-interaction can remain too low to drive complete exploration (snapshots 2 and 5). Terracette-like patterns emerge only in an intermediate window (snapshots 3 and 6). The red square marks the parameter pair used in other simulations.}
\label{FigS4}
\end{figure*}

\revisionthree{In Figure \ref{FigS4} we sweep over parameter combinations corresponding to different recovery and revisit timescales, performing individual simulations. Each simulation uses $N=1$ to avoid the confounding effects of multiple interacting agents. 
To identify periodic surface banding, we compute the autocorrelation of slope-aligned elevation transects and detect the first prominent peak, which indicates the dominant tread spacing. We select an autocorrelation threshold of $\bar{R}(\lambda) > 0.75$ to indicate terracette-like bands (see Supplementary Information Section D for calculation details). 
When $1/\mu<s/l$, resources recover before agents revisit them. We ignore such values, as this would prevent interaction via the resource map $\mathcal{R}$, critical for driving agent exploration. When $1/\mu \gg s/l$, agents may over-deplete the landscape and constrain motion, preventing trail reinforcement. Terracette formation is only observed in a narrow intermediate regime where recovery is neither too fast nor too slow relative to revisit (snapshots 3 and 6 in Figure \ref{FigS4}).}

\subsection*{Evaluating terrain anisotropy}

\begin{figure}[h!]
\includegraphics[width=0.45\textwidth]{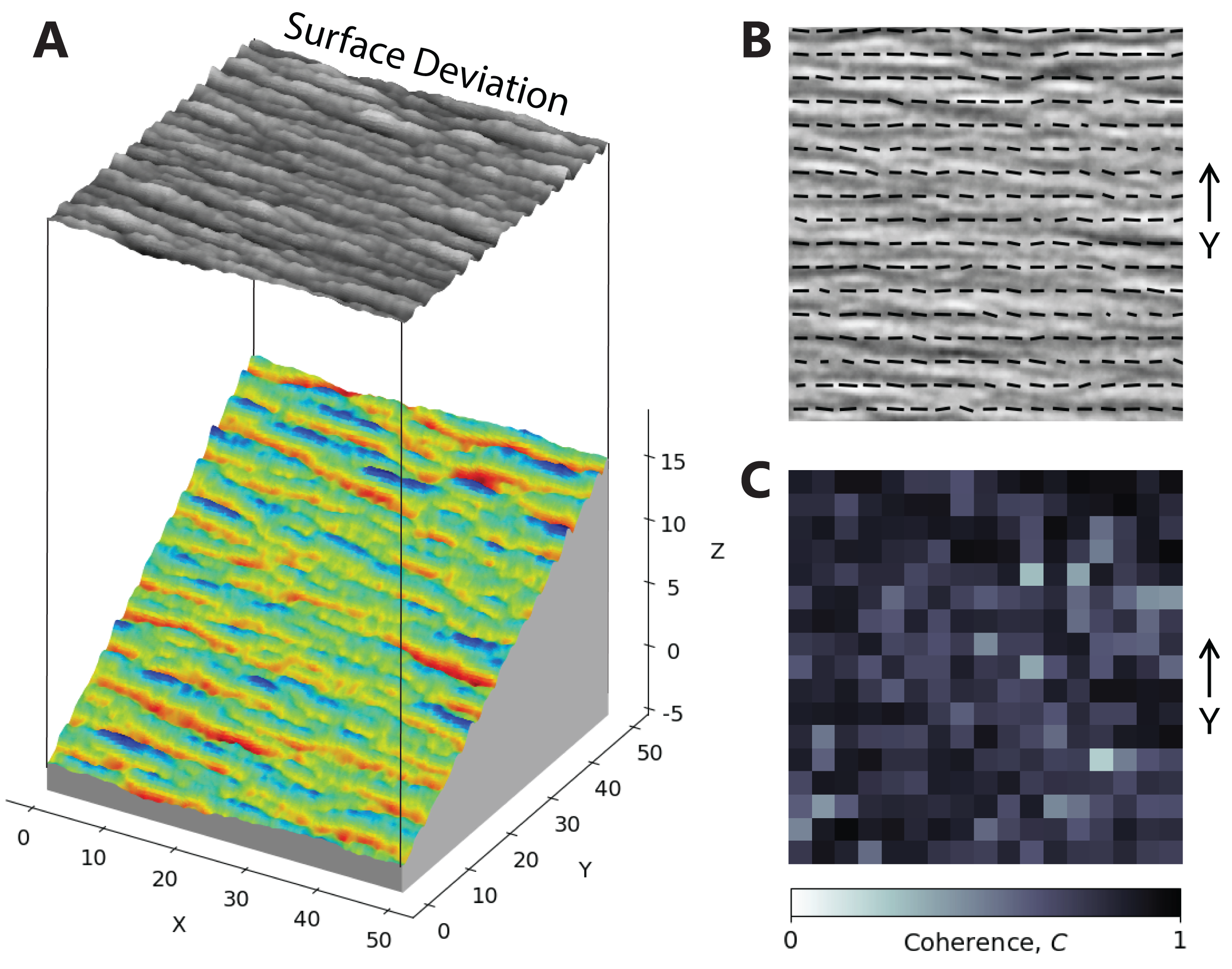}
\caption{\textbf{Coherence analysis.} (A) The major slope is subtracted to reveal surface deviation, $\Delta \mathcal{H}$. (B) The orientation field of the surface deviation, showing a highly ordered surface. The needles represent the estimated ridge-valley orientations within each sampling window. (C) The coherence field, illustrating the coherence of gradient vectors within each window in panel B. For details on the computation of the orientation field, see the Supplementary Information.}
\label{fig:coherence_analysis} 
\end{figure}

\revisionthree{While the autocorrelation analysis identifies periodic band formation, it is most effective when patterns are unidirectional and regularly spaced. Prior methods, including direct measurements of risers and treads~\cite{walsh2001physical}, path-geometry statistics~\cite{jia2023distribution}, and 2-D Fourier analysis of satellite imagery~\cite{hellman2020detection}, likewise assume well-aligned or regularly spaced features. A more general measure of terrain anisotropy is therefore required to quantify terracette structure across irregular landscapes.}
Here, we adopt a directional‑field approach originally developed for edge tracking~\cite{kass1987analyzing} and widely used in fingerprint analysis~\cite{bazen2002systematic}. To isolate small-scale surface features, large-scale trends must be removed. We do this simply by subtracting the initial terrain from the final heightmap, yielding a flattened deviation map $\Delta \mathcal{H} = \mathcal{H}(t=t_f) - \mathcal{H}(t=0)$ (Figure~\ref{fig:coherence_analysis}A). This deviation map is then partitioned into square windows of side length $w$. For a square terrain of size $s$, the number of non‑overlapping windows is $M=\left(s/w\right)^2$. 
Within each window, we compute the local height gradients $G_x$ and $G_y$ and their second-order moments $G_{xx}=\langle G_x^2\rangle$, $G_{yy}=\langle G_y^2\rangle$, and $G_{xy}=\langle G_xG_y\rangle$. These moments are then used to define the local coherence for that window,

\vspace{-10pt}
\begin{equation}
    C = \frac{\sqrt{\left(G_{xx}-G_{yy}\right)^2 + 4G_{xy}^2}}{G_{xx}+G_{yy}}
\end{equation}

where $C=0$ indicates isotropy, and therefore no preferred orientation, whereas $C=1$ denotes perfect alignment of ridges and valleys (see Figures~\ref{fig:coherence_analysis}B and~\ref{fig:coherence_analysis}C). The window‑level coherence when averaged over all $M$ windows yields a global order parameter,

\vspace{-10pt}
\begin{equation}
 \bar{C}=\frac{1}{M}\sum_{m=1}^M C_m
\end{equation}

Because terracettes form regularly spaced, directionally consistent steps, well-developed ordered patterns drive the surface toward $\bar{C}\rightarrow 1$. Notably, the choice of window size $w$ can influence the resulting values. In particular, coherence remains above zero even for $\beta=0$ and $\sigma=0^\circ$, since local structure is always present even at small window scales. A comparison of different window sizes is provided in Figure~S5. 

\begin{figure}[hb!]
\includegraphics[width=0.48\textwidth]{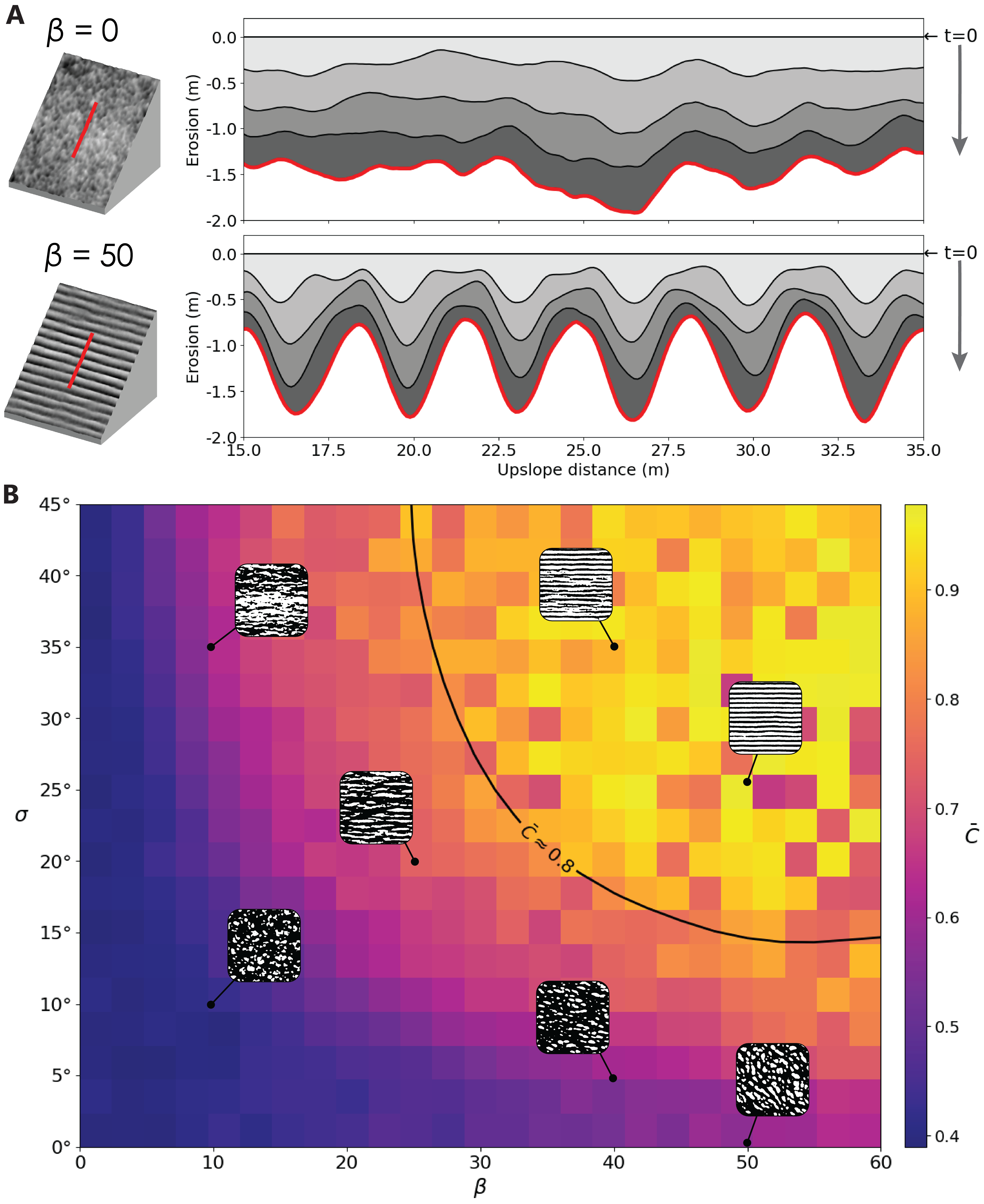}
\caption{\textbf{Terracette formation is governed by energetic sensitivity and cost of travel.} (A) Evolution of surface deviations extracted along identical upslope transects ($\sigma = 40^\circ$). The major slope has been subtracted to reveal the surface deviation at multiple time points throughout the simulation. Low $\beta$ yields diffuse, unstructured erosion, while high $\beta$ produces reinforced periodic bands of erosion. (B) Mean coherence $\bar{C}$ across $\sigma$ and $\beta$ parameter space. Anisotropy increases with both slope and sensitivity, reflecting stronger energetic costs and sensitivity. Selected $\Delta \mathcal{H}$ maps are thresholded at their mean erosion to highlight emergent patterns, with darker regions indicating troughs or trails. Terracette-like features begin to emerge once coherence exceeds $\bar{C} \gtrsim 0.8$.}
\label{fig:data} 
\end{figure}

\subsection*{Influence of energetic sensitivity and cost of travel}
Temporal erosion profiles show that an agent's energetic sensitivity, $\beta$, strongly modulates pattern formation (Figure~\ref{fig:data}A). At low $\beta$, agents perform a nearly uniform random walk, so erosion remains diffuse regardless of terrain slope. At high $\beta$, however, erosion localizes and assembles into periodic bands parallel to the contour lines of the slope. To disentangle the respective roles of slope angle $\left(\sigma\right)$ and energetic sensitivity $\left(\beta\right)$, we plot a heatmap of the order parameter $\bar{C}$ on the $\left(\sigma,\beta\right)$ plane (Figure~\ref{fig:data}B). We find that coherence increases monotonically with both variables. To further illustrate the gradual transition from scattered erosion to coherent bands, we overlay simulated erosion maps on the heatmap; each map is thresholded at the mean depth, so darker tones indicate greater erosion, and distinct terracette‑like motifs emerge once $\bar{C}\gtrsim0.8$. Again using autocorrelation of gradient-direction transects, we obtain a mean tread spacing of $\lambda \approx 3.38\,\text{m}$ across the terracette-motif regime (Figure~S6). 

\begin{figure}[hb!]
\includegraphics[width=0.45\textwidth]{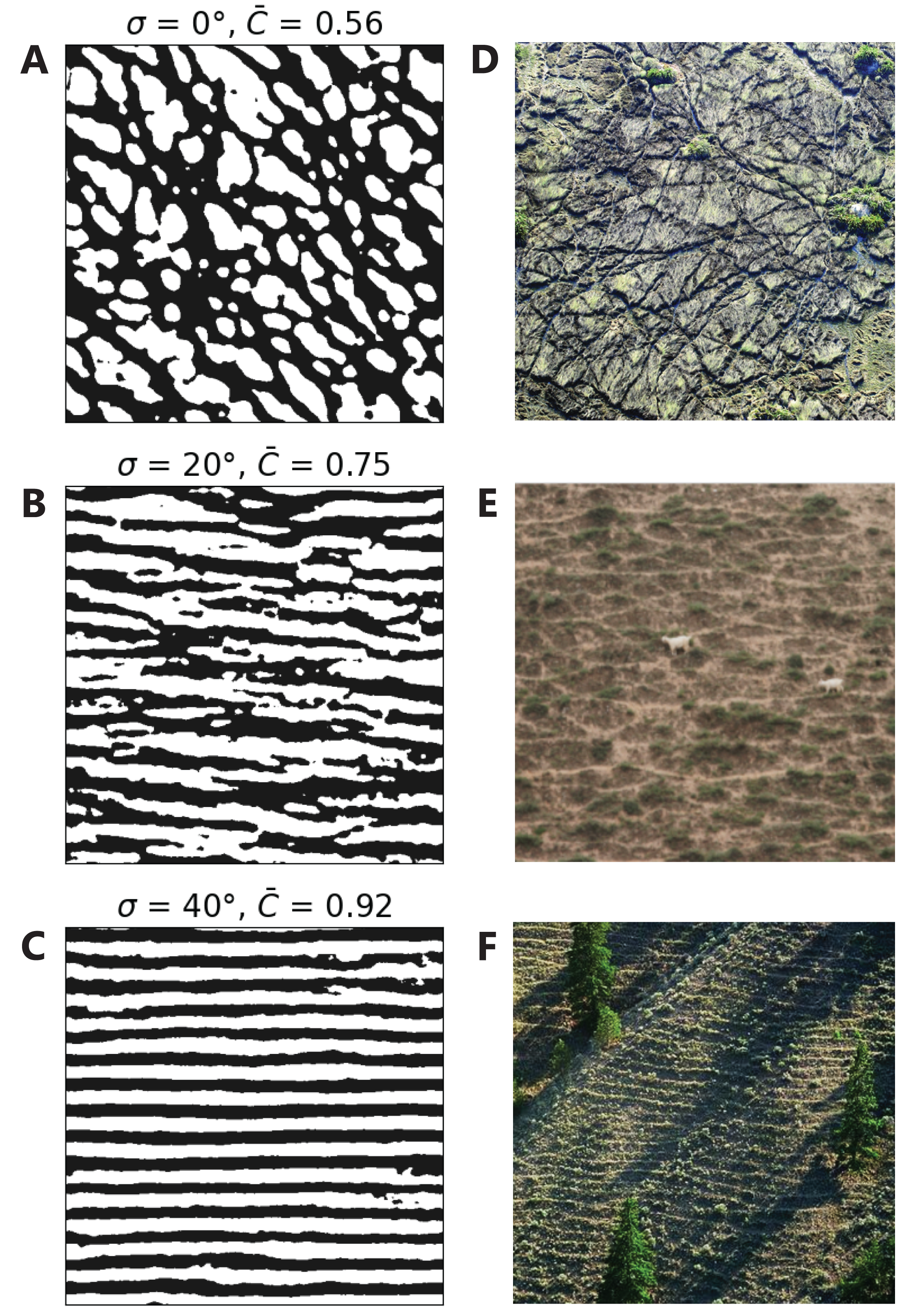}
\caption{\textbf{Pattern formation in real settings.} 
(A-C) Thresholded $\Delta \mathcal{H}$ maps (50\,m\,$\times$\,50\,m) shown from an orthophotographic perspective illustrate erosion patterns across increasing slope angles. As slope increases, a transition from isotropic trails to anisotropic terracettes occurs. 
(D - F) Real-world erosion patterns that resemble the modeled results in (A), (B), and (C).
(D) Marshland with animal trails in the Okavango Delta, Africa (from \cite[Figure~162]{propper2015future}).
(E) Goat paths on the Loess Plateau near Lanzhou, China (from \cite[Figure~1]{jin2019goat}).
(F) Cow paths near Spences Bridge, British Columbia (from \cite{druterracettes}).}
\label{fig:qualitative_panels} 
\end{figure}

When energetic sensitivity is high, discernible paths form on all slopes, though their coherence varies. Figure~\ref{fig:qualitative_panels}A-C show simulated erosion maps alongside real-world snapshots in Figures~\ref{fig:qualitative_panels}D-F, together illustrating the consistent progression from winding tracks to well‑ordered terracettes. On flat ground $\left(\sigma=0^{\circ}\right)$, erosion yields a loose, meandering network with a modest coherence of $\bar{C}=0.56$. At a moderate slope of $\sigma=20^{\circ}$, those paths remain interconnected but align more, raising coherence to $\bar{C}=0.75$. Once the slope reaches $\sigma=40^{\circ}$, erosion self‑organizes into regularly spaced, parallel bands, the characteristic tread‑and‑riser pattern of terracettes, driving coherence up to $\bar{C} = 0.92$. The close visual agreement between simulations and real terrain underscores how increasing slope and energetic sensitivity jointly transform diffuse wandering into highly ordered terracette structures.

\section*{Discussion}
Despite the longstanding discourse on terracettes, empirical studies have been limited. To our knowledge, only one field experiment on a natural hillslope has explicitly examined terracette formation~\cite{higgins1982}, while most subsequent work remains purely observational. Field experiments are difficult due to the long timescales and logistical challenges. Computational approaches, on the other hand, offer an easier means of probing underlying mechanisms.
Except for one early simulation that coupled vegetation growth with mass wasting~\cite{gallart1993computer}, no mechanistic model has yet explained how terracettes form, whether by geophysical processes or by grazing ungulates. Here we introduce an agent‑based framework in which simulated ungulates, modeled as Boltzmann active walkers, graze and trample an erodible slope. When locomotion incurs a slope‑dependent energetic cost, the model shows that terracette‑like bands emerge solely from repeated trampling and reciprocal feedback with the substrate, without explicit communication among agents or auxiliary geomorphic processes. 

The proposed mechanism is inherently stigmergic: every footstep compacts soil and depletes forage, reshaping the local energy landscape and biasing subsequent movement. Paths that run parallel to contour lines are progressively reinforced until periodic tread‑and‑riser bands appear. This trade‑off between locomotion cost and forage intake echoes field observations, where cattle and goats favor cross‑slope routes and terracette treads to minimize vertical work~\cite{ganskopp2000least,jin2016livestock,jin2019goat, reichman1981mammal}. Our simulations reproduce these trends, showing that steeper slopes amplify vertical costs, steering agents laterally and raising overall terrain coherence, while larger $\beta$ values strengthen that bias and accelerate terracette-like band formation (Figure~\ref{fig:data}). We also find a characteristic tread spacing $\lambda\approx 3.38\,\mathrm{m}$. This value is consistent with spacing set by partial overlap of foraging paths ($\lambda \sim 2r_{\mathcal{D}}$), in line with the geometric grazing hypothesis of~\cite{howard1987dimensions}.


Despite its successes, the framework rests on several simplifying assumptions and has a bounded regime of applicability. Terracettes can emerge even from a single agent via self-interaction, while increasing the number of agents typically accelerates erosion and the rate of pattern formation (Figure~S2). At sufficiently high densities, however, collective grazing can exhaust and homogenize the resource layer, eliminating the gradients required for stigmergic alignment and robust terracette formation. 
$N=3$ on \revisionthree{the} 50\,m$^2$ domain should be interpreted as an intensely grazed patch that is revisited many times over, rather than as a real-time measure of grazing impacts. In practice, we set an intermediate, artificial regrowth rate $\mu$ to sustain agent-resource feedback \revisionthree{(Figure \ref{FigS4})}. 
\revisionthree{These modeling choices effectively compress ecological timescales, such that simulation time steps should not be interpreted as direct predictions of real-world temporal dynamics.} Additionally, the use of a linear regrowth parameter $\mu$ oversimplifies vegetation growth and ignores trampling-induced soil degradation and delayed vegetation recovery under heavy grazing~\cite{trimble1995cow,bilotta2007impacts}. Accordingly, $\mathcal{R}$ is best interpreted as a stigmergic ``memory'' of recent traffic rather than a mechanistic vegetation model. 


More broadly, the framework treats erosion as compaction alone, omitting shear-induced creep and density changes that harden substrates under repeated trampling~\cite{romero2023modelling}. \revisionthree{An artifact of this simplification} is that the terrain remains unrealistically ``soft'', as can be seen in the pronounced paths of Figure~\ref{fig:qualitative_panels}A. In nature, ungulates rarely create such discernible paths on flat terrain ($\sigma\approx 0^{\circ}$) except on especially soft substrates, such as marshland (Figure~\ref{fig:qualitative_panels}D), or in patchily grazed landscapes~\cite{walker1986effect}. \revisiontwo{Because no process exists to redistribute material, simulations can \revisionthree{also} produce runaway erosion and near-vertical risers that, in reality, would exceed the angle of repose and make them prone to failure and other geophysical mass-wasting processes \cite{hartwig2025gully}. 
To avoid runaway erosion, we elected to terminate simulations when the riser height exceeds a prescribed threshold (1.5\,m), restricting our attention to the pre-failure regime.} 
\revisionthree{To assess whether the emergence of terracette structure depends on this stopping criterion, we implemented an alternative termination strategy in which compaction gradually saturates with erosion depth (ignoring creep), allowing simulations to converge without a strict cutoff (Figure~S7). Across matched initial conditions, this depth-limited formulation produced similar terracette morphology and global coherence, while preventing runaway erosion. These results suggest that model terracette formation is also robust to a convergence-based termination criterion.}

\revisiontwo{In nature, terracette patterns have been observed to change through time~\cite{higgins1982}, with existing steps degrading as new ones form. This suggests that terracettes exist in quasi-equilibrium, with biogenic and geophysical forces acting in tandem. By construction, our model does not attempt to capture this long-term balance.} As such, the landscape features reproduced by the model do not represent equilibrium patterns but rather transient ones. 
\revisionthree{The smooth initial terrain considered here provides a controlled setting in which agents can generate and respond to slopes created through their own interactions. As mentioned, however, terracette formation in nature may also occur on previously degraded or irregular terrain. 
To examine this alternative, we initialized simulations with spatially correlated noise added to the base terrain. Across a range of noise amplitudes, agents still formed coherent quasi-parallel terracette patterns, although the placement of paths was visibly influenced by the underlying features of the initial terrain. (Figure~S6). Across noise levels, coherence exhibited a modest decline, but cross-slope banded structure remained evident (Figure~S6).} 


It is \revisiontwo{also} worth noting that slope usage is not species agnostic. Larger animals inherently pay higher energetic penalties, and beyond that, some slopes may be infeasible depending on biomechanics. For instance, goats have been seen traversing steeper interconnecting paths than horses and cattle~\cite{howard1987dimensions,taylor1972running}, while elephants may avoid some steep terrain altogether~\cite{wall2006elephants}. Such variation could be captured in the model by adjusting energetic sensitivity ($\beta$), tuning the ratio $\omega_v/\omega_m$, or imposing strict slope cutoffs as in~\cite{gilks2009mountain}. \revisiontwo{In the present model, we also assume symmetric costs for uphill and downhill travel. In reality, both movement costs and effective reach are asymmetric on sloped terrain, and depend on species-specific biomechanics. Implementing different costs for uphill and downhill movement and direction-dependent foraging radii could bring the model closer to species-specific grazing behavior.}

These considerations also highlight the need for empirical grounding. Rigorous validation will therefore require applying our coherence metric to high‑resolution topography from GPS or LiDAR and comparing simulated $\bar{C}$ values with field-collected data. Likewise, animal‑borne GPS, accelerometry, and heart‑rate loggers~\cite{brosh2006energy,dickinson2021behaviour} could quantify path lengths, turning angles, and energetic costs on real slopes—data essential for calibrating path characteristics and energetic sensitivity $\beta$. 
\revisiontwo{Additionally, empirical estimates of the timescales over which terracettes form, persist, and degrade would provide the grounding needed to relate simulation steps to real landscape evolution.} While the spatial and temporal scales of terracette formation may still impede field data collection, biomimetic robots~\cite{Kalogroulis2024GoatHoof} may offer a testbed for future experiments.

Alongside empirical validation, theoretical extensions may further broaden the framework. A natural next step would be to couple the present optimal foraging mechanism to explicit geomorphic transport laws and vegetation dynamics. Such a model would yield a more complete biogeomorphic framework, capable of linking measurable soil-transport properties and grazing pressures to long-term terracette dynamics.
Interestingly, the erosion maps in Figure~\ref{fig:qualitative_panels}A-C resemble classic Turing patterns~\cite{cross1993pattern,ball2015forging}. Future work could attempt to recast terracette formation in a reaction‑diffusion framework, as has been done for vegetation bands on hillslopes~\cite{ge2023hidden, deblauwe2011environmental} and \revisiontwo{other} grazing patterns \cite{ge2022foraging}. Such a continuum model could test whether interactions between erosion and foraging decisions can be captured by activator–inhibitor dynamics. 

\revisiontwo{To conclude,} our simplified model cannot rule out the possibility that geophysical mechanisms contribute to natural terracette formation. Indeed, shear stresses and resultant soil creep generated by grazer trampling may accelerate \revisiontwo{trail formation on sloped terrains or periodically erase steps as risers approach failure.} Nevertheless, our work demonstrates that large-scale terracette geometry can emerge in silico from local movement rules that trade energy against forage on sloped terrain. More broadly, it underscores how decentralized agents interacting with a dynamic energy landscape self-organize into persistent, long-range-ordered networks, a motif that recurs in active matter systems~\cite{hallatschek2023proliferating}, vascular and branching architectures \cite{tero2010rules,ben1994generic}, and the formation of ant and pedestrian trails~\cite{sumpter2003nonlinearity, helbing1997active, couzin2003self, gilks2009mountain}. By embedding terracettes within this shared biophysical framework, we offer both a mechanistic explanation for a classic geomorphic debate and a bridge to comparative studies of path formation across scales and taxa.


%

\begin{acknowledgments}
We thank Dr. Ahmed El Hady for organizing the 2024 Konstanz School of Collective Behavior, where this project was conceptualized, and all participants for their early feedback and discussion. We thank the University of Washington Libraries for providing access to print materials essential to this study.
\textbf{Funding:} S.B. acknowledges funding from NSF CAREER IOS-1941933 and Schmidt Sciences, LLC.
\textbf{Author contributions:}
Conceptualization, methodology, writing—original draft: B.S.
Writing—review and editing: B.S., A.C., S.B. Software: B.S., L.G. 
Visualization: B.S., A.C., L.G. 
Supervision, funding acquisition: S.B.
\textbf{Competing interests:} The authors declare no competing interests.
\textbf{Data and materials availability:} All Python code to run simulations, generate figures, and reproduce results is publicly available on \href{https://github.com/bhamla-lab/terracettes-active-walkers}{GitHub}.
\end{acknowledgments}

\end{document}


\title{Supplementary Materials for\\
Moo-ving mountains: grazing agents drive terracette formation on steep hillslopes}

\author{Benjamin Seleb}
\affiliation{Interdisciplinary Graduate Program in Quantitative Biosciences, Georgia Institute of Technology, Atlanta, GA, United States}

\author{Louis Gonz\'alez}
\affiliation{School of Chemical and Biomolecular Engineering, Georgia Institute of Technology, Atlanta, GA, United States}
\affiliation{Department of Chemical and Biological Engineering, University of Colorado, Boulder, CO, United States}

\author{Atanu Chatterjee}
\affiliation{School of Chemical and Biomolecular Engineering, Georgia Institute of Technology, Atlanta, GA, United States}
\affiliation{Department of Chemical and Biological Engineering, University of Colorado, Boulder, CO, United States}

\author{Saad Bhamla}
\email{saadb@chbe.gatech.edu}
\affiliation{School of Chemical and Biomolecular Engineering, Georgia Institute of Technology, Atlanta, GA, United States}
\affiliation{Department of Chemical and Biological Engineering, University of Colorado, Boulder, CO, United States}

\maketitle

\vspace{1cm}

This PDF file includes:
\begin{itemize}
  \item Supplementary Text
    \begin{enumerate}[label=\Alph*., leftmargin=*, nosep]
      \item Landscape Modification
      \item Simulation Parameters
      \item Coherence Calculation
      \item Autocorrelation (Spacing) 
      Calculation
      \item Noisy Initial Landscapes
      \item Depth-limited Compaction
      
    \end{enumerate}
  \item Figures S1 to S7
  \item Table S1

\end{itemize}

\vspace{0.5cm}
Other Supplementary Materials for this manuscript include the following:
\begin{itemize}
  \item Movies S1 to S2
\end{itemize}

\clearpage

\section*{Supplementary Text}

\subsection{Landscape Modification}

Agents alter their landscape through two mechanisms at each footstep: resource depletion and terrain erosion. Both are modeled using radially symmetric landscaping kernels centered on each footstep location (Figure~\ref{FigS1}).

\begin{figure}[h]
\centering
\includegraphics[width=0.98\textwidth]{figures/SIfigures/FigS1-01.png}
\caption{
(A) The two landscaping functions used in the model, depicted as functions of $r$ with parameters labeled. Both functions are shown in their positive form and are subtracted to produce erosion and resource depletion at each footstep.  
(B) Surface cross-sections showing the effect of a single footstep on sloped terrain with increasing pitch: $\sigma = 0^\circ$, $20^\circ$, and $40^\circ$. The terrain profile is shown as a solid line, and the shaded region represents the resulting depression from a single application of the erosion kernel $\mathcal{F}(x,y)$.  
Here, $\alpha = 1$\,m and $r_\mathcal{D} = 2$\,m are used purposefully to enhance visualization.
}

\label{FigS1}
\end{figure}

\subsubsection{Updating the Resource Map}

At each footstep $(x_k, y_k)$, resources are depleted within a radius $r_D$ using a uniform depletion kernel $\mathcal{D}(x, y)$:

\[
\mathcal{D}(x, y) =
\begin{cases}
\gamma, & r \leq r_D \\
0, & \text{otherwise}
\end{cases}
\quad \text{where } r = \sqrt{(x - x_k)^2 + (y - y_k)^2}
\]

In our simulations, $\gamma = 1$, indicating complete depletion within the radius $r_D$. The resource map is updated as:

\[
\mathcal{R}_{t+1}(x, y) = \left[\mathcal{R}_t(x, y) + \mu\right] - \sum_{a=1}^{N} \sum_{k=1}^{n} \mathcal{D}(x_k, y_k)
\]

We select a replenishment rate $\mu$ that allows for substantial resource loss to still exist after an agent has made approximately 2-3 loops of the periodic boundary condition. To prevent values from exceeding physical bounds, $\mathcal{R}$ is clipped to the interval $[0, 1]$:

\[
\mathcal{R}_{t+1}(x,y) \leftarrow \max \left\{ 0, \min \left\{ 1, \mathcal{R}_{t+1}(x,y) \right\} \right\}
\]

This means, with a realistic depletion radius (i.e., neck length), they may follow the edge of a depleted path, rather than spacing themselves $r_D$ from a depleted path as expected. To account for this, we set $r_D$ to twice the intended sensing range to prevent agents hugging edges of fully depleted patches.

\subsubsection{Updating the Heightmap}

Terrain erosion is modeled using a smooth, circular footprint kernel $\mathcal{F}(x, y)$, which tapers toward the edges. To simulate natural variation in foot placement, each footstep location $(x_k, y_k)$ is perturbed by random offsets:

\[
(\widetilde{x}_k, \widetilde{y}_k) = (x_k + v_x, y_k + v_y)
\quad \text{with } v_x, v_y \in [-\eta, \eta]
\]

The erosion kernel is defined as:

\[
\mathcal{F}(x, y) =
\begin{cases}
\alpha \cos \left( \dfrac{\pi r'}{2 r_F} \right), & r' \leq r_F \\
0, & \text{otherwise}
\end{cases}
\quad \text{where } r' = \sqrt{(x - \widetilde{x}_k)^2 + (y - \widetilde{y}_k)^2}
\]

The terrain is updated by subtracting all footprints from the heightmap:

\[
\mathcal{H}(x, y, t+1) = \mathcal{H}(x, y, t) - \sum_{a=1}^{N} \sum_{k=1}^{n} \mathcal{F}(\widetilde{x}_k, \widetilde{y}_k)
\]

Figure~\ref{FigS1} shows both the resource depletion kernel $\mathcal{D}$ and erosion kernel $\mathcal{F}$ as a function of $r$, along with an example of how a single footprint depression appears on slopes of varying steepness. The uniform kernel $\mathcal{D}$ reflects the assumption that grazers consume resources within their accessible radius with no preference, while the bell-shaped erosion kernel $\mathcal{F}$ represents a basic pressure gradient beneath the agent.

\subsection{\label{sec:parameters}Simulation Parameters}

Unless noted otherwise, simulations use the baseline values listed in Table~S1.
Two parameters of practical importance are the \texttt{simulation resolution} and the \texttt{erosion threshold}. The \texttt{simulation resolution} defines the horizontal grid size used to discretize the terrain surface, agent movement, and environmental modifications. Higher resolutions allow for finer spatial detail but increase computational cost. Note that each grid cell's elevation is treated as a continuous variable.

To determine when simulations should terminate, we define an \texttt{erosion threshold} based on the maximum terrain deviation. Instead of tracking raw elevation values, we compute the surface deviation at each time step $t_i$, represented by the map $\Delta \mathcal{H}(x, y,t=t_i)$ by subtracting the initial heightmap from the current heightmap:

\[
\Delta \mathcal{H}(x, y,t=t_i) = \mathcal{H}(x, y, t = t_i) - \mathcal{H}(x, y, t = 0)
\]

As described in the main text, this yields a \emph{flattened} deviation map with the major slope removed. We then compute the maximum height difference (amplitude) of the remaining surface deviations:
\[
A(t) \;=\; \max_{x,y}\{\Delta \mathcal{H}(x,y,t)\}\;-\;\min_{x,y}\{\Delta \mathcal{H}(x,y,t)\}.
\]
Simulations terminate once $A(t)$ exceeds the predefined \texttt{erosion threshold}.

Simulations support the use of multiple agents, as they interact indirectly via the heightmap and resource map. To examine the effects of agent density, we varied the number of agents $N\in\{1,3,5\}$ and quantified resultant terrain morphology and the anisotropy order parameter $\bar C$ (SI section C). As can be seen in Figure~\ref{FigS2}, while a single agent is still able to generate high-order landscape features, increasing $N$ accelerates erosion/pattern formation and reduces runtime (terminated once $A(t)$ meets the \texttt{erosion threshold}). On our reference workstation (AMD Ryzen 9 7950X, 64\,GB RAM; CPU implementation), a typical $N{=}3$ simulation ($<10000$ timesteps) can be completed in $\sim$ 3 minutes using the provided scripts. All Python scripts are available via \href{https://github.com/bhamla-lab/terracettes-active-walkers}{GitHub}.

\begin{figure}[h]
\centering
\includegraphics[width=0.98\textwidth]{figures/SIfigures/FigS2-01.png}
\caption{
Effect of agent number on coherence and runtime. (A) Across slopes $\sigma\in\{0^\circ,20^\circ,40^\circ\}$ and $\beta = 50$, terracette coherence $\bar C$ remains high for sloped terrain and is largely insensitive to the number of agents ($N=1,3,5$), indicating that even a single agent can generate order above the $\sigma=0^\circ$ baseline. (B) Time to termination (depth threshold) decreases with $N$, consistent with faster erosion/pattern formation under higher trampling rates. Each box summarizes 30 independent runs per $(N,\sigma)$ condition. The central line is the median, the box spans the interquartile range (Q1–Q3), whiskers extend to the most extreme points within $1.5\times\mathrm{IQR}$ of the quartiles, and points beyond the whiskers are plotted as outliers. Colors encode slope angle $\sigma$. 
}
\label{FigS2}
\end{figure}

The per-meter movement cost in the vertical and horizontal directions is defined by $\omega_v$ and $\omega_h$, respectively. These costs directly influence the energetic bias on agents...  contributing to the cumulative cost of motion $\varepsilon$. Before exponentiation, we scale this value to the range $[0,1]$ using min-max normalization to produce $\tilde{\varepsilon}(\phi_i)$. This step is completed to avoid numerical underflow when exponentiating the scaled cost $\rho(\phi_i)$. Because of this step, the absolute values of $\omega_v$ and $\omega_h$ become largely irrelevant. Instead what matters is the \emph{ratio} of per-meter costs: 
\[
\omega_v/\omega_h.
\]
In Figure~\ref{FigS3}, we sweep $\omega_v/\omega_h$ for two values of $\beta$ (all other parameters as in Table~S1) and report the coherence $\bar C$ for individual runs (points). Two visible trends can be observed. First, increasing $\omega_v/\omega_h$ (i.e., penalizing vertical motion more strongly than horizontal) increases the resultant coherence of simulations. This is consistent with a stronger lateral bias that strongly favors horizontal trajectories. Second, reducing $\beta$ lowers the coherence of all simulations, irrespective of the per-meter costs. For the primary simulations, we select a ratio $\omega_v/\omega_h$ (solid vertical line) that is likely to generate coherence values above the visible terracette threshold at $\bar C\approx0.8$.

\begin{figure}[h]
\centering
\includegraphics[width=0.9\textwidth]{figures/SIfigures/FigS3-01.png}
\caption{
Coherence vs. vertical–horizontal cost ratio. Individual simulations (points) show the dependence of the coherence $\bar C$ on the cost ratio $\omega_v/\omega_h$ for two energetic sensitivities ($\beta=50$ and $\beta=25$). Slope angle $\sigma = 40^\circ$. 
Variation arises from the underlying stochasticity of the model, with solid curves depicting a rolling average. Increasing $\omega_v/\omega_h$ biases motion laterally and raises $\bar C$, whereas $\beta$ controls the overall sensitivity irrespective of ratio. The horizontal dashed line marks the approximate threshold ($\bar C\approx0.8$) above which terracette-like banding becomes visually apparent, and the vertical line indicates the cost ratio used in the primary simulations, which lies well above this threshold for both $\beta$ values.  
}
\label{FigS3}
\end{figure}

All parameters not under exploration were fixed at their baseline values (Table~S1) to isolate their effects. Our aim is to test a minimal mechanistic model by which energetic bias and stigmergic feedback yield terracette-like patterns. Accordingly, several quantities are used in representative (not empirically measured) ranges, and the baseline parameterization lies within the terracette-forming windows identified.

\subsection{Coherence Calculation}

To evaluate the anisotropy of terrain erosion, we compute a coherence metric based on directional field analysis. This method quantifies the degree to which surface features (e.g., ridges and valleys) are aligned.

We begin by computing the final terrain deviation map:

\[
\Delta \mathcal{H}(x, y) = \mathcal{H}(x, y, t = t_f) - \mathcal{H}(x, y, t = 0)
\]

We then calculate the gradients:

\[
G_x = \frac{\partial \Delta \mathcal{H}}{\partial x}, \quad G_y = \frac{\partial \Delta \mathcal{H}}{\partial y}
\]

These gradients are divided into square windows of size $w$, and second-order moments are computed within each window:

\[
G_{xx} = \langle G_x^2 \rangle, \quad
G_{yy} = \langle G_y^2 \rangle, \quad
G_{xy} = \langle G_x G_y \rangle
\]

From these, the coherence is calculated as:

\[
C = \frac{\sqrt{(G_{xx} - G_{yy})^2 + 4G_{xy}^2}}{G_{xx} + G_{yy}}
\]

Local coherence values range from 0 (no preferred direction) to 1 (perfect alignment). The global coherence $\bar{C}$ is computed by averaging over all windows. To visualize directional structure, we also compute the orientation angle $\theta$ in each window:

\[
\theta = \arctan(V_j, V_i) \mod \pi
\]

Where:

\[
V_i = \frac{1}{2}(G_{xx} - G_{yy}) - \frac{1}{2} \sqrt{(G_{xx} - G_{yy})^2 + 4G_{xy}^2}, \quad
V_j = G_{xy}
\]

This angle represents the dominant orientation of surface features within each window. Figure \ref{fig:coherence} illustrates this process on a flat terrain (where ridge/valley direction is easiest to see) and additionally shows how the choice of window size $w$ influences the measured coherence.

\begin{figure}[h]
\includegraphics[width=0.8\textwidth]{figures/SIfigures/FigS4-01.png}
\caption{\label{fig:coherence} Coherence analysis. (A) Surface deviation computed from a flat terrain. We select a flat terrain to give an alternative visualization of coherence/orientation fields to that in the report. (B) Comparison of window size $w$ and resultant coherence values for a number of simulations with varying $\sigma$. Each ensemble contains 10 simulations. Different values of $w$ were used in computing the coherence of the same group of simulations (5 angles, 50 simulations total). 
Increasing $w$ lowers the minimum coherence value, though coherence changes minimally  above $\sigma = 30^\circ$. Interestingly $\sigma = 30^\circ$ has the greatest variance in coherence. (C) The direction field. (D) The computed coherence for each window. (E) The orientation field, depicted the major direction of valleys and ridges.}
\end{figure}

\subsection{Autocorrelation (Spacing) Calculation}

To estimate terracette spacing we sample the final deviation map
$\Delta\mathcal{H}(x,y)=\mathcal{H}(x,y,t_f)-\mathcal{H}(x,y,0)$
along $P$ slope–parallel transects oriented in the gradient direction (perpendicular to contours). On our inclined plane these are straight lines. Let the $p$-th transect be the sequence of grid points
$(x_p,y_p)$, and define the 1-D transect
\[
z_p \;=\; \Delta\mathcal{H}\!\bigl(x_p,y_p\bigr).
\]
For each transect we compute the autocorrelation
$R_p$ at each discrete lag $k\in\{0,\dots,K\}$. This is implemented using \texttt{statsmodels.tsa.stattools.acf} with \texttt{nlags}$=K$.)
We then average across all $P$ transects to obtain a robust estimate,
\[
\bar R(k) \;=\; \frac{1}{P}\sum_{p=1}^P R_p(k),
\]
and identify the first nonzero peak $\bar{R}(\lambda)$ where $\lambda$ is the dominant periodicity (terracette spacing in meters) using \texttt{scipy.signal.find\_peaks}.\\
In practice we use $P=5$ evenly spaced transects and set $K$ large enough to
cover the spacings of interest. Figure~\ref{fig:autocorrelation}C displays the calculated spacing $\lambda$ for multiple simulations across $\sigma$ and $\beta$ parameter space. These values are computed on the same simulations used to generate Figure~7B. We visualize spacing confidence by scaling marker size with $\bar{R}(\lambda)$. Notably, we find a high-confidence region that coincides with the coherence threshold $\bar{C}\!\approx\!0.8$ described in the main text. Taking the average of $\lambda$ within this region, we find a mean terracette spacing of 3.38\,m. \\
Similar to the coherence, we can also use $\bar{R}(\lambda)$ as a morphology indicator (e.g., classifying periodic surface features when $\bar{R}(\lambda)$ exceeds a specified threshold in Figure~6). However, this metric only works as a diagnostic on landscapes with a single dominant slope.

\begin{figure}[h]
\centering
\includegraphics[width=0.98\textwidth]{figures/SIfigures/FigS5-01.png}
\caption{\label{fig:autocorrelation}Autocorrelation-based measurement of terracette spacing and its relation to coherence.
(A) Low-slope example (blue outline, $\beta=45$, $\sigma=2.5^\circ$). The deviation map with slope-parallel transects depicted as colored lines. The 1-D profiles are extracted along these transects, from which the lagged autocorrelation $R(k)$ of each profile is computed. Correlations decay without a clear periodic peak.
(B) High-slope, terracetted example (red outline, $\beta=50$, $\sigma=35^\circ$). Transects are nearly sinusoidal and $R$ shows repeated peaks, with $k = \lambda$ being the lag at the first nonzero peak (red dot). 
(C) Simulation sweep over $(\beta,\sigma)$ space, summarizing terracette spacing. Point color encodes the estimated spacing $\lambda$, and point area encodes the peak autocorrelation $\bar{R}(\lambda)$. The gray curve marks the same $\bar C\!\approx\!0.8$ contour from the coherence map in Figure~7B, delineating the terracette-forming regime. Simulation values from which panels (A) and (B) are derived are shown using the same outline colors.}
\end{figure}

\red{\subsection{Noisy Initial Landscapes}
To examine terracette formation on irregular terrain, we extended the base landscape initialization by adding spatially correlated noise to the sloped heightmap. To introduce terrain irregularities, we generated a spatially correlated noise field using Perlin noise and added it to this base slope:
\begin{equation*}
\mathcal{H}\left(x,y,t=0\right)=y\tan\sigma + \nu(x,y)
\end{equation*}

where $\nu(x,y)$ is a bounded noise field scaled by a noise amplitude parameter. Perlin noise was generated using the Python \texttt{noise} package. To maintain compatibility with the periodic boundary conditions used in the simulations, we generated tileable noise fields by enforcing periodicity in both spatial directions. Specifically, we used the \texttt{pnoise2} function with periodic repetition (easily recreatable using the Python scripts available via \href{https://github.com/bhamla-lab/terracettes-active-walkers}{GitHub}).
The resulting noise field was normalized to the range $[0,1]$ and scaled by a noise amplitude parameter.

\begin{figure}[h!]
\centering
\includegraphics[width=\textwidth]{figures/SIfigures/FigS6-01.png}
\caption{Terracette formation on initially noisy landscapes. (A) Example initial terrains generated by adding spatially correlated noise to a baseline sloped surface, shown for noise amplitudes of 0 m, 2 m, and 5 m. (B) Resulting terrains after agent-driven erosion for the corresponding initial conditions in Panel (A). Simulations continue to produce cross-slope banded structures, though patterns become visually less uniform at higher noise amplitudes. (C) Terrain coherence ($\bar{C}$) as a function of noise amplitude for noise scales 0–5 m. Box-and-whisker plots show results from 10 replicate simulations per condition. A modest decrease in coherence is observed with increasing terrain roughness. (D) Simulation durations (timesteps prior to termination) across noise amplitudes remain comparable across conditions. All simulations were performed using identical parameters ($\beta=50$, $\sigma=40^\circ$, $N=3$ agents). The same tileable noise kernel (fixed seed) was scaled to produce different noise amplitudes across conditions.
    }
\label{fig:FigS6}
\end{figure}

Simulations were performed across a range of noise amplitudes to assess the sensitivity of terracette formation to initial terrain irregularity (Figure~S6). Increasing noise amplitude produced progressively more irregular initial landscapes, which in turn influenced the uniformity of emergent paths. Nevertheless, cross-slope banded structures remained visible across all tested conditions. To quantify this effect, we computed terrain coherence across 10 replicate simulations for each noise amplitude (Figure~S6C). While coherence exhibited a modest decline with increasing terrain roughness, values remained relatively high across the tested range, indicating that anisotropic structure formation persists even in irregular landscapes.
}

\red{\subsection{Depth-limited Compaction}

To test whether the simulation results depend on the choice of stopping criterion, we replace the hard erosion cutoff used in the main text with a physically motivated soil-compaction saturation term,
\begin{equation*}
    g(d) = \mathrm{max}\left( 0, 1 - \left( \frac{d}{d_{\rm sat}} \right)^K \right)
\end{equation*}
where $d$ is the local erosion depth (the difference between the initial and current terrain height at each grid cell) and $d_{\rm sat} = 25$ grid units is the saturation depth beyond which the soil can no longer compact. This function multiplies the erosion kernel at each footstep:
\begin{equation*}
    \mathcal{H}(x, y, t+1) = \mathcal{H}(x, y, t) - \sum_{a=1}^{N} \sum_{k=1}^{n} \mathcal{F}(\widetilde{x}_k, \widetilde{y}_k) \cdot g(d).
\end{equation*}
When the ground is fresh ($d = 0$), $g = 1$ and erosion proceeds at full strength; as a groove deepens toward $d_{\rm sat}$, $g$ decays steeply to zero, naturally preventing runaway compaction. The exponent of 6 ensures that $g$ remains near unity throughout the pattern-forming regime and only activates close to $d_{\rm sat}$, preserving the positive-feedback dynamics that drive terracette formation.

To ensure that $g$ remains near unity throughout the pattern-forming regime and only activates close to $d_{\rm sat}$, we use $K = 6$. This preserves the positive-feedback dynamics that drive terracette formation. Both simulations use identical initial conditions, parameters, and random seed ($\sigma = 40^{\circ}$, $\beta = 50$, and $K = 6$). The results of the simulations are shown in Figure~\ref{fig:FigS7}.

\begin{figure}[h]
\centering
\includegraphics[width=0.98\textwidth]{figures/SIfigures/FigS7-01.png}
\caption{ Robustness test of depth-limited compaction. Panels (A) and (B) are simulation snapshots at termination for both the hard-cutoff and self-converged simulations. Morphology is consistent with deeper colors, demonstrating that a longer simulation time $t$ before convergence is met yields deeper troughs. This is shown in Panel (C). Panel (D) shows that the unconstrained erosion leads to the terrain approaching the cutoff threshold, whereas the addition of the $g(d)$ term causes the terrain depth to saturate, but never reach, the hard threshold. At this steady state, the standard deviation between each successive terrain state becomes smaller than 0.05. Global coherence $C$ is roughly the same (0.955 versus 0.959 for original and the $g(d)$ versions, respectively), indicating no morphological change occurs when applying this non-arbitrary simulation terminus. The simulations have a minimum run time of $t = 3000$ steps. 
    }
\label{fig:FigS7}
\end{figure}

Erosion maps from the original hard-cutoff simulation (Panel A; $ \bar{C} = 0.955$, $t = 4473$) and the $g(d)$ simulation with no hard cutoff (Panel B; $ \bar{C} = 0.959$, $t = 5281$) show indistinguishable terracette banding, spacing, and orientation. Upslope cross-sections along the midline (Panel C) confirm that the periodic tread-and-riser wavelength and phase are preserved, with the $g(d)$ troughs slightly deeper owing to the additional timesteps before self-convergence. The time evolution of the maximum terrain deviation (Panel D) illustrates the key distinction: the original simulation grows until it is terminated externally at the imposed threshold (gray dotted line), whereas the $g(d)$ simulation follows the same trajectory but levels off naturally as compaction saturates, providing a self-consistent stopping mechanism without the need for an arbitrary cutoff.}

\clearpage


\begin{sidewaystable}[h]
\caption{\label{tab:table1}Summary of model parameters. \textit{All quantities and units are intended to approximate real-world scales but do not correspond to physical measurements. An asterisk (}*\textit{) indicates the value used in main text simulations and figures.}}
\begin{ruledtabular}
\begin{tabular}{lllll}
\textrm{Parameter}&
\textrm{Unit}&
\textrm{Symbol}&
\textrm{Values explored}&
\textrm{Reasoning}\\
\texttt{Number of agents} & - & \textit{N} & 3 & See supplemental text and Figure~S2\\
\texttt{Terrain size} & m & \textit{s} & 50& Baseline\\
\texttt{Major slope} & degrees & $\sigma$ & 0-45& Encompassing slope range\\
\texttt{Path length} & m & $l$ & 10& See supplemental text and Figure~S4\\
\texttt{Step size} & m & $\delta$ & 0.5& Approximate ungulate stride length\\
\texttt{Number of visual bins} & - & $B$ & 36& Reasonably high-resolution visual acuity (10$^\circ$)\\
\texttt{Energetic sensitivity} & arbitrary & $\beta$ & 0-60& Baseline 0. Swept until terracettes regularly observed\\
\texttt{Maintenance cost} & $\text{J\,m}^{-1}$ & $\omega_h$ & 0.95& See supplemental text and Figure~S3\\
\texttt{Vertical cost} & $\text{J\,m}^{-1}$ & $\omega_v$ & 8.25& See supplemental text and Figure~S3\\
\texttt{Trampling depth} & m & $\alpha$ & 0.01& Small compaction to ensure gradual erosion\\
\texttt{Footprint radius} & m & $r_\mathcal{F}$ & 1& Basic unit length of the spherical cow\\
\texttt{Footstep noise magnitude} & m & $\eta$ & 0.5& Small noise ($\le\delta$) for foot placement variability\\
\texttt{Resource depletion strength} & proportion & $\gamma$ & 1& Agents set to fully deplete resources for simplicity\\
\texttt{Resource regrowth rate} & proportion/timestep & $\mu$ & 0.1& See supplemental text and Figure~S4\\
\texttt{Depletion radius} & m & $r_\mathcal{D}$ & 2& Baseline reach length ($> r_\mathcal{F}$)\\
\texttt{Coherence window size} & m & $w$ & 0.5, 1, 1.5, 2*& See supplemental text and Figure~S5\\
\texttt{Simulation resolution} & m & - & 0.1& Good balance between resolution and computation\\
\texttt{Erosion threshold} & m & - & 1.5& Cutoff to prevent unrealistic terrain deformation\\
\end{tabular}
\end{ruledtabular}
\end{sidewaystable}

\clearpage

\noindent MOVIE S1. Simulation timelapse of a flat terrain ($\sigma = 0^\circ$) with $\beta = 50$. See file \texttt{Movie\_S1.mp4}.

\vspace{1em}

\noindent MOVIE S2. Simulation timelapse of a sloped terrain ($\sigma = 35^\circ$) with $\beta = 50$, showing the evolution of periodic erosion bands reminiscent of terracettes. See file \texttt{Movie\_S2.mp4}.

\newpage